\documentclass[aps,pre,amssymb,twocolumn,showpacs]{revtex4}
\usepackage{graphicx}
\usepackage{epsfig}
\usepackage{subfigure}
\usepackage{graphics}
\usepackage{setspace,bm}
\begin{document}

\renewcommand{\bottomfraction}{0.8}
\setcounter{bottomnumber}{3}

\title{Rheology of carbon nanotube dispersions}
\author{Y. Y.~Huang, S. V.~Ahir and E.M.~Terentjev}

\affiliation{Cavendish Laboratory, University of Cambridge, J.J.
Thomson Avenue, Cambridge CB3 OHE, U.K. }

\begin{abstract}
\noindent We report on rheological properties of a dispersion of
multi-walled carbon nanotubes in a viscous polymer matrix.
Particular attention is paid to the process of nanotubes mixing
and dispersion, which we monitor by the rheological signature of
the composite. The response of the composite as a function of the
dispersion mixing time and conditions indicates that a critical
mixing time $t^*$ needs to be exceeded to achieve satisfactory
dispersion of aggregates, this time being a function of nanotube
concentration and the mixing shear stress. At shorter times of
shear mixing, $t<t^*$, we find a number of non-equilibrium
features characteristic of colloidal glass and jamming of
clusters. A thoroughly dispersed nanocomposite, at $t>t^*$, has
several universal rheological features; at nanotube concentration
above a characteristic value $n_c \sim$2-3 wt\% the effective
elastic gel network is formed, while the low-concentration
composite remains a viscous liquid. We use this rheological
approach to determine the effects of aging and re-aggregation.
\end{abstract}



\maketitle

\section{Introduction}\label{Intro}

The pursuit of well dispersed nanotubes into a given matrix is a
fundamental problem that still hinders research and development a
long time since they were brought to global
attention~\cite{Iijima1991}. Monitoring the quality of dispersion
within a given system gives rise to additional problems. While
clustering of spherical particles has been studied well, for both
spherical and highly asymmetrical (platelets, rods and
fibers)~\cite{Persello1994,Hobbie1998,Fry2002,
Lin-Gibson2003,Mohraz2004}, there are no reliable direct
techniques of observing carbon nanotubes in the bulk of a
composite suspension. All optical methods cut off below a length
scale of $\sim$0.2-0.5$\mu$m; all electron microscopy methods (so
prominent in observations of individual nanotubes) can only
provide information about the sample surface, i.e. only
representative for the selected fields of view. This leaves
reciprocal space techniques and, more importantly, global indirect
techniques of characterizing the dispersed nanocomposites; each of
these techniques suffers from the unavoidable difficulty in
interpretation of results. A recent review gives a summary of such
approaches, their strong and weak aspects, and
prospects~\cite{Nalwa2005}.

If efficient and economically viable bulk processing of
nanotube--polymer composites is to be realized, a well developed
understanding of responses to simple steady-state shear flow is
required. In this paper we concentrate on the analysis and
interpretation of rheological characteristics of nanocomposites at
different stages of their dispersion and subsequent aging (tube
re-aggregation). Some rheological data has appeared in the recent
literature~\cite{Potschke2002,Hobbie2004,Du2004,Potschke2004} but
to our understanding, no work has yet been undertaken to apply
rheological data to characterize the state of dispersion directly,
and moreover, to investigate the effect of conditions and mixing
time on the quality of nanotube dispersion.

In itself, dispersion is a spatial property whereby the individual
components (in this case nanotubes) are spread with the roughly
uniform number density throughout the continuous supporting
matrix. The first challenge is to separate the tubes from their
initial aggregated assemblies, which is usually achieved by local
shear forces. However, a homogeneous suspension ideally achieved
after mixing is not necessarily a stable state: the removal of a
shearing force may open the way to re-aggregation. At very low
concentrations, the conditions of an ideal-gas occur, when the
dispersed objects do not interact with each other. However, for
nanotubes with very high persistence length, the Onsager treatment
of anisotropic suspensions~\cite{deGennes93book} suggests that the
crossover concentration when the rod-like objects start
interacting and significantly biasing their orientational pair
correlation can be very low indeed~\cite{Islam2004,Song2005}.
Experimental evidence also suggests external hydrodynamic forces
can also induce clustering in highly anisotropic
suspensions~\cite{Schmid2000flow}.  There are several classical
ways of surface treatment, which improves the colloidal stability
of nanotubes~\cite{Nalwa2005}; in all cases it is more challenging
than in usual sterically stabilized colloids because of the
unusual depth of the primary van der Waals minimum due to the high
polarizability of nanotubes. Many authors have suggested that when
the loading of nanotubes is above a critical value, a network
structure can form in the nanocomposite system during
mixing~\cite{Potschke2002,Du2004,Shaffer1998Disp}. Elastic gel,
arising from such an entangled nanotube
network~\cite{Kinloch2002,Kharchenko2004}, may prevent individual
tube motion and thus serve as an alternative mechanism of
stabilization.

In this work we choose to work with untreated tubes, since our
main goal is to examine the dispersion and re-aggregation
mechanisms. We believe that the state of dispersion in a given
composite dispersion can be ascertained by measuring the global
rheological properties of the system. The viscosity of the mixture
has a direct correlation with the spatial and orientational
distribution of nanotubes in the matrix. This can be used as a
physical signal with which to monitor the quality of dispersion,
as long as the interpretation of the rheological signal is
calibrated. By studying the rheology of a viscous polymer mixed
with nanotubes at different stages of dispersion and aging we aim
to provide, for instance, an answer to the question of how long
one should shear-mix their nanotube-polymer sample to achieve a
suitable level of dispersion. Our conclusions are somewhat
surprising: the required mixing time is so long and the required
mixing shear so high that one might question the quality of
nanotube dispersion of many famous experiments in the last decade.

The discussion of this work is split into three parts. Firstly,
after giving the details of sample preparation and measuring
procedures, we give an overview on how the composite viscosities
are affected by tube concentration and mixing time. Secondly, we
develop a general interpretation of the frequency dependence of
rheological data, and possible dispersion states have been
correspondingly deduced, especially focusing on higher
concentrations and elastic gel network formation. Finally, the
stability of the dispersion state is discussed in the last part of
this paper. In Conclusions, we cast a critical look at our own
work as well as the claims of nanotube dispersion in the
literature.

\section{Experimental}

\begin{figure}[b]
\centering
\resizebox{0.33\textwidth}{!}{\includegraphics{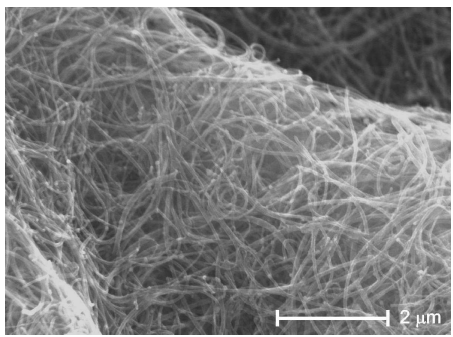}}
\caption{A typical Scanning Electron Microscopy (SEM) image of
nanotube agglomerates from the supplier, showing the entangled
nature of raw samples prepared by the CVD method. } \label{bundle}
\end{figure}

Multi-walled carbon nanotubes (Nanostructured \& Amorphous
Materials, Inc.) are used with purity verified as $>$95\%. These
nanotubes, prepared by the method of Catalytic Vapor Deposition,
are found to be tangled in agglomerates, Fig.~\ref{bundle}, in
contrast to many other sources producing tubes in densely packed
and aligned bundles obtained by the CVD technique. The
manufacturer specified dimensions, tube length $L \sim 5-15\,\mu$m
and outer diameter $d\sim 60-100$\,nm, are confirmed by direct SEM
observation, also indicating the tube persistence length $l_p\sim
0.5-1\,\mu$m. The pristine nanotubes were lightly ground by pestle
and mortar prior to usage without surface-modification at any time
during processing.

The physics of semiflexible ``worm-like chains'' (representing our
tubes) in a solvent (our matrix) is
well-developed~\cite{Colby03book}. Given the parameters of
diameter $d$, persistence length $l_p$ and total arc length $L$,
we can estimate the characteristic overlap concentration, which
theoretically marks the boundary between dilute (individual tubes
in solution) and semidilute (interpenetrating tubes) regimes.
Explaining this in detail is not a task for us and we refer the
interested reader to the polymer physics literature. The model
result for the overlap volume fraction $\phi_c$ depends on whether
one accounts for the so-called excluded volume interaction (i.e.
prohibits tubes from crossing each other). It is often neglected
in the first approximation, but we shall specifically need this
excluded volume interaction to account for the elastic entangled
network emerging in our system. \footnote{Potential energy of
excluded volume interactions, $U\sim kT n^2 d^3 R^3 \approx kT d
L^2/R^3$ is balanced against the entropic free energy $F \sim kT
R^2/(l_pL)$ of the semiflexible chain, giving the chain size
$R_g\sim (l_pd)^{1/5}L^{3/5}$. The overlap is then at a volume
fraction $\phi_c \sim d^2L/R_g^3$.} Therefore, the overlap volume
fraction is estimated as $\phi_c \sim d^{7/5}l_p^{-3/5}L^{-4/5}$.
For our tube parameters this gives the volume fraction $\phi_c
\sim 0.003-0.008$.
 However, to
make comparisons with our experiments (in which we measure the
loading by the weight percent), we need to convert the volume
fraction $\phi = V_{\rm tube}/V_{\rm total}$ into the weight
fraction. Using the density of nanotubes, $\rho_{\rm tube} \sim
2.1\, \hbox{g/cm}^3$ (from the manufacturer data sheet), we
estimate the overlap to occur at $n_c \sim 0.5-1.5$\,wt\%. Above
this concentration, the semidilute solution of self-avoiding
chains becomes increasingly  entangled and develops the elastic
modulus~\cite{Colby03book}.

The polymer matrix used throughout this work is PDMS
(Polydimethylsiloxane) Sylgard 184$^{TM}$ from Dow Corning. We
have not crosslinked the elastomer networks with the usual Sylgard
curing agent, instead using only the PDMS compound as a
well-characterized viscoelastic liquid. The compound has a
molecular weight, $M_{w}\sim 18.5kD$, density $\rho \sim 1.2\,
\hbox{g/cm}^3$ and the apparent viscosity of 5.6\,Pa.s at
25$^{\circ}$C.

\begin{figure} 
\centering \resizebox{0.4\textwidth}{!}{\includegraphics{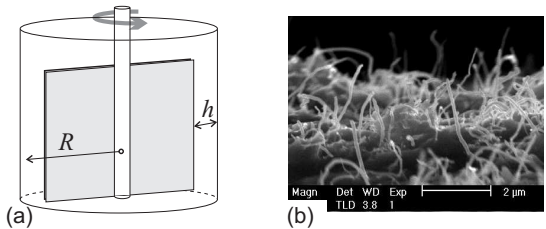}}
\caption{(a) Scheme of the mixer container, with the relevant
dimensions labelled for calculation of shear stress. \ (b) An SEM
image of freeze-fractured surface of a well-dispersed 7wt\%
composite ($t_{\rm mix}=61\,$h).} \label{mixer}
\end{figure}

Batches of samples (each of about 2g total weight) were prepared
with different weight fractions: 0.5wt\%, 1wt\%, 2wt\%, 4wt\% and
7wt\%, by direct addition of the PDMS compound onto the dry
nanotubes. The samples were dispersed by using an Ika Labortechnik
mixer; the geometry of shear mixing is illustrated in
Fig.~\ref{mixer}(a), specifying the key dimensions used in the
analysis below. Throughout the dispersion process, the rotation
speed of the paddle was kept at 1000 rpm and the mixing
temperatures kept at 30$\pm$ 0.5$^{\circ}$C. The high speed of
mixing and comparably low mixing temperature should ensure that
only stimulated dispersion and little spontaneous re-aggregation
to take place during mixing (cf. ~\cite{Martin2004}).

After a desired length of mixing time, aliquots of the sample were
removed from the mixer and their rheological characteristics were
measured immediately, as well as after different standing times
(the period of time elapsed after the mixer has been switched off
and before rheological tests were conducted). The standing samples
were stored at room temperature.

According to the geometry of the mixer, the shear stress applied
to the nanotube-PDMS mixture can be approximated as $\sigma \sim
\eta R \omega /h$, where $R \approx 7$mm and $h \approx 1.5$mm are
the radius and the gap indicated in the diagram, $\omega \approx
1.2 \cdot 10^5 \hbox{rad.s}^{-1}$ the angular frequency of the
paddle and $\eta$ is the viscosity of PDMS. The resulting estimate
of shear stress is of the order of $1$~MPa, increasing with the
increasing nanotube loading. The ultimate tensile stress of sigle-
and multi-walled nanotubes is in the range of
200-900~MPa~\cite{Schadler1998, Salvetat1999Shery,Walters1999},
therefore, direct scission of the tubes is unlikely to occur
during mixing.

Monitoring the quality of nanotube dispersion in a continuous
polymer matrix is a perennial problem, with very few experimental
techniques available to resolve it. Electron microscopy, which is
the only method offering real-space resolution on the scale of
nanotubes, is an inherently surface technique. Attempting to
dissolve or ion-etch the polymer to reveal the tubes, immediately
leads to their re-aggregation. Making samples very thin to allow
transmission microscopy makes nanotubes interact with surfaces
much more than with the bulk. Figure~\ref{mixer}(b) shows the SEM
image of a freeze-fractured interface of a well-mixed 7wt\%
polymer composite. It suggests the tubes are homogeneously
dispersed, but is far from offering any proof of dispersion
quality. The main point of our present work is to develop and
alternative (rheological) quantitative method of monitoring the
dispersion.


Rheological measurements were staged on a stress-controlled
Rheometrics DSR (dynamic stress rheometer) connected to a water
bath heater to ensure a consistent temperature of 30$\pm$
0.5$^{\circ}$C. A cone-and-plate geometry (25 mm diameter, 0.1 rad
cone angle, gap 0.01 mm) was utilized and a stress-controlled (set
at 100~Pa) frequency sweep experiment was performed to monitor the
linear viscoelastic response of each sample. In order to compare
the characteristic viscosity of each composite, we chose the
frequency of 50Hz.

For this cone-plate shear geometry we define the Reynolds number
by $Re\approx \dot{\gamma} z^{2}\rho/\eta$, with $\rho$ the PDMS
mass density, $\dot{\gamma}$ the strain rate and $z$ the average
distance between the test plates. We find $Re \leq 10^{-3}$ for
all strain rates used throughout our experiments.

We ensured (by comparing the data at different shear rates) that
the rheometer provides oscillatory shear measurements of linear
response, with the real and imaginary viscosities
($\eta'(\omega)$~\&~$\eta''(\omega)$ respectively). The equivalent
information contained in the storage and loss shear moduli
(($G'(\omega)$~\&~$G''(\omega)$) is useful for identifying gel
properties and jamming of clusters.

\section{Results and Discussion}

\subsection{Mixing time of dispersion}\label{vistime}

\begin{figure} [b]
\centering \resizebox{0.4\textwidth}{!}{\includegraphics{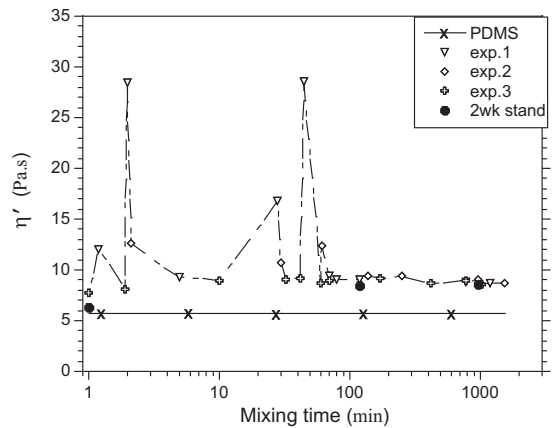}}
\caption{Plot of the real viscosity $\eta'$, sampled at fixed
frequency 50Hz, against the time of mixing for the 1wt\%
nanotube-PDMS sample. Three different experiments have been
carried out and the results shown on the same plot. The data shows
that for $t_{\rm mix}<~100$min, the viscosity reading is erratic;
mixing for longer periods leads to reproducible well-dispersed
composite. Black dots show the selected samples tested after 2
weeks standing.  } \label{threeruns}
\end{figure}

In order to determine the effect of mixing time on the degree of
nanotube dispersion, three identical experiments were performed
for samples with 1wt\% concentration of nanotubes, with the
results shown in Fig.~\ref{threeruns}. Each test was conducted on
an aliquot of the composite after a certain time of continuous
mixing of a sample; this was repeated for three separate mixtures.
The rheological test protocol involves taking each sample through
a full frequency sweep and measuring its shear response; the plot
only shows the value of $\eta'$ sampled at one frequency. The
values of the viscosity obtained for the short mixing times
($t<100$min) have erratic values, such that no trend can be
assigned to the viscosity variation with increasing mixing time.
We shall discuss the origins of this effect at greater detail
below. Presently, it is important to note that after a certain
time of mixing, these erratic values turn to a consistent value of
composite viscosity, which is the same in different experiments
and not much affected by further mixing. We interpret this
characteristic time, $t^*$, as the minimal time required to
achieve the complete dispersion at the given concentration of
tubes and the mixing shear stress.

\begin{figure} 
\centering \resizebox{0.4\textwidth}{!}{\includegraphics{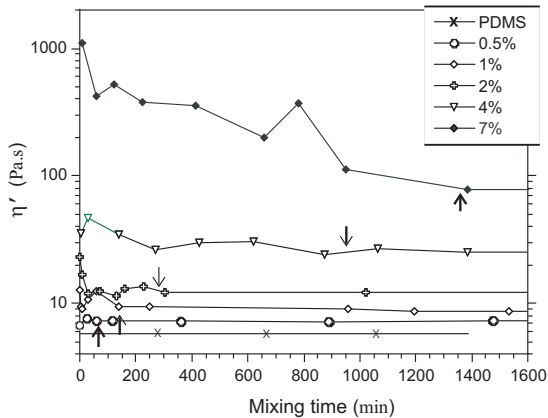}}
\caption{Plot of the real viscosity, $\eta'$ at 50Hz, against the
time of mixing, for a range of different weight fractions of
nanotubes in the composite. The viscosity axis is logarithmic to
allow all samples to fit on the same plot, since the viscosity
increase with tube concentration is significant, while the time
axis is linear to allow low concentrations to be resolved. The
arrows mark the critical time $t^*$ for each concentration.}
\label{all}
\end{figure}

The comparison of different experiments for a 1wt\% sample in
Fig.~\ref{threeruns} is a good illustration of universality.
Overall, such change in viscosity with mixing time occurs for all
mixtures investigated, Fig.~\ref{all}. Note that the axes scale in
Fig.~\ref{threeruns} is linear-log (to capture the whole time
range and emphasize the variations in $\eta'$), while the
log-linear axes are used in Fig.~\ref{all} (to focus on the early
times and allow all viscosity data on the same plot). In
Fig.~\ref{all}, for each sample, the data points at very long
times are not shown (e.g. $t_{\rm mix}=61$h for 7wt\%) since no
further change in $\eta'$ was observed.

The key features of the dispersion process, the large increase in
overall viscosity with nanotube loading, the erratic values at
short mixing times and the consistent reproducible viscosity
reading for each composite, dispersed for $t> t^*$, are reproduced
for each sample. Although we do not show the corresponding plots,
it was found that this characteristic time for the onset of the
stable region (i.e the signature of the complete dispersion) was
independent of the sampling frequency and is equally evident true
for both $\eta'$ and $\eta''$.

In general, erratic behavior for short-time-mixed (i.e.
incompletely dispersed) composites implies the nanotube clusters
have a large variation in sizes and concentration across the bulk
mixture. Each batch of sample can have its own dispersion
characteristic after being mixed for a short time. This results in
an unavoidably unpredictable results of our rheological
measurements.

\begin{figure}
\centering
\resizebox{0.37\textwidth}{!}{\includegraphics{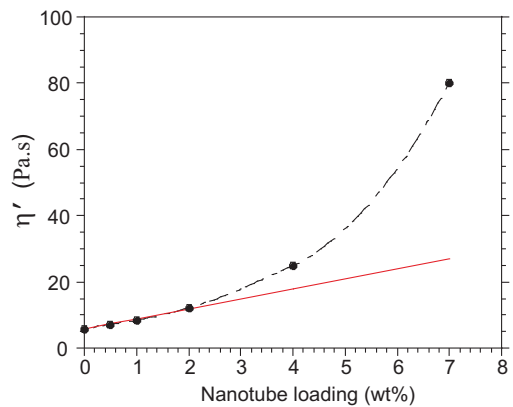}}
\caption{Plot of the dispersion plateau viscosity (at 50Hz)
against the weight fraction of nanotubes in the composite. The
solid line shows the initial linear (non-interacting) regime; the
dashed line is a guide to an eye in the non-linear (entangled)
regime.} \label{finalvisc}
\end{figure}

\begin{figure} 
\centering
\resizebox{0.37\textwidth}{!}{\includegraphics{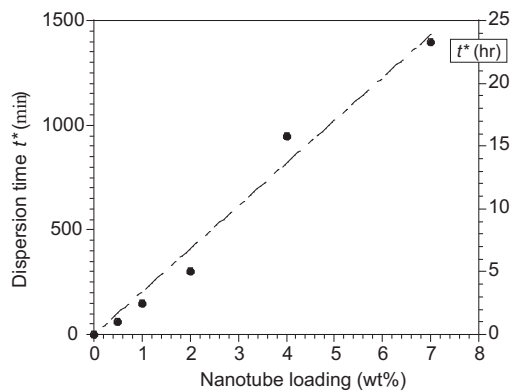}}
\caption{Plot of the critical time for dispersion, $t^*$ obtained
from data in Fig.\ref{all}, against the weight fraction of
nanotubes. The right axis shows the same time in hours. The dashed
line is the linear fit $t^* \propto n$ (see text).} \label{time}
\end{figure}

Clearly, every batch of the same concentration should tend to the
same dispersion state after being mixed long enough, $t>t^*$. The
resulting dispersion should consist of a fairly uniform
distribution of nanotube, which does not change with further
mixing and yields a specific and reproducible rheological response
as indicated by the plateau regimes in Fig.~\ref{all}. Two
important results can be further extracted from this data: the
dependence of the final plateau viscosity, and of the critical
mixing time $t^*$, on the tube concentration. These are presented
in Figs.~\ref{finalvisc} and \ref{time}. The concentration
dependence of the final plateau value of the dispersion viscosity
$\eta_{s}'$ shows the expected linear increase at low
concentrations, corresponding to the non-interacting Einstein like
suspension viscosity; the line in Fig.~\ref{finalvisc} is $\eta' =
\eta'_{\rm PDMS}(1+103.5\, n)$, where $n$ is the volume fraction
of nanotubes calculated from the value of wt\% in the
plots.\footnote{There is no point comparing the slope of
$(1+103.5\, n)$ with, e.g., a classical result for the dispersion
of rigid rods: our tubes are anything but rods, cf.
Fig.~\ref{bundle}, and the more appropriate models of dilute
semiflexible polymer solutions are too ambiguous to offer a
reliable number.} At concentrations above $n^*\sim$2wt\% the
deviations from the linear regime become noticeable indicating the
tube interactions and eventually entanglements in the dispersed
composite.

Figure~\ref{time} shows a fairly linear correlation between the
critical mixing time $t^*$ and tube concentration. The following
simple argument suggests that this is an expected feature: Let us
define $dE (n,t)$ to be the elementary energy needed to disperse
one nanotube in a composite with a given environment characterized
by the volume fraction $n$ and mixing time $t$. Similarly, let
$P(n,t)$ to be the power transmitted during the shear mixing,
where $t$ is the time of mixing ($t<t^*$). Thus
 $$
\int^{n}_{0}dE(n,t)=\int^{t}_{0}P(n,t)dt .
 $$
To simplify this crude analysis, we define $\overline{E} (n)$ as
the average constant energy needed to disperse a nanotube, and
similarly $\overline{P} (n)$ as the average power. This will lead
to $n \overline{E} (n) = \overline{P} (n) t^*$ and, therefore, $
t^* = n [\overline{E} (n)/\overline{P} (n)]. $ Since physically,
both $dE(n)$ and $P(n)$ describe the same process, it is not
surprising that their ratio is nearly constant and the primary
$n$-dependence is linear, as indeed is seen in Fig.~\ref{time}.

\subsection{Dispersion rheology}

\subsubsection{Clusters at $t_{\rm mix}<t^*$}

\begin{figure} [b]
\centering \resizebox{0.4\textwidth}{!}{\includegraphics{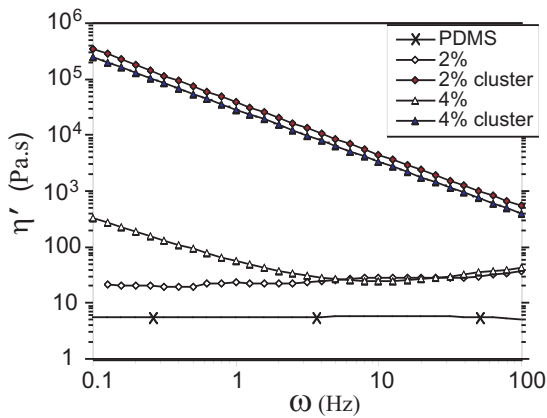}}
\caption{Frequency sweep of viscosity for 2wt\% and 4wt\% samples
mixed for $t_{\rm mix} < t^*$. The bold symbols show the results
when the sample aliquot in the rheometer was found to contain
visible nanotube clusters: note the extremely high values of
(arrested) $\eta'$, the same for different samples. }
\label{clustersA}
\end{figure}

\begin{figure} 
\centering
\resizebox{0.32\textwidth}{!}{\includegraphics{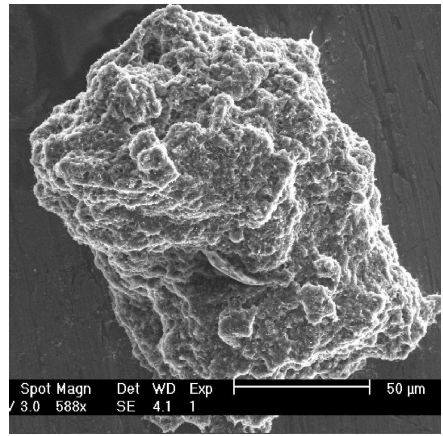}}
\caption{SEM image of a typical compact nanotube cluster found in
a sample mixed only for a few minutes.} \label{clustersB}
\end{figure}

It has already been stated in the previous section that the
rheological response for $t_{\rm mix}<t^*$ is erratic. This
requires some detailed discussion. Occasionally, in such poorly
dispersed samples, $\eta'$ is obtained several orders of magnitude
greater than what would otherwise be expected (see
Fig.~\ref{threeruns}). In Fig.~\ref{clustersA} we show two such
measurements, as a frequency sweep, for a 2wt\% (mixed for 2min)
and a 4wt\% (mixed for 1 hour) samples each.  In each sample, one
of the measurements is a meaningful representation of a
complex-fluid response. The other (very high) reading is identical
for different samples and reflects an arrest of the rheometer
plates when a nanotube cluster is wedged between them. Such poorly
mixed composites are invariably found to contain visible nanotube
clusters, which are the main reason for the accidentally high
viscosity reading. When such clusters are so large as to block the
rheometer plates (minimal cone-and-plate gap is ~$0.01$mm), the
measurement is obviously flawed. The scanning electron microscope
(SEM) image of one typical cluster in Fig.~\ref{clustersB} shows a
size of ~$0.05mm$ and a much more compacted structure in
comparison with the initial entangled tube agglomerates
(Fig~\ref{bundle}). When such plate blocking occurs, the viscosity
readings are obviously unaffected by the nanotube concentration.
Longer mixing-times ensures the clusters are broken down into a
homogeneous dispersion, hence no further erratic behavior is
observed.

\subsubsection{Well dispersed state, $t_{\rm mix}>t^*$}\label{sec:wellmix}

\begin{figure} 
\centering \resizebox{0.4\textwidth}{!}{\includegraphics{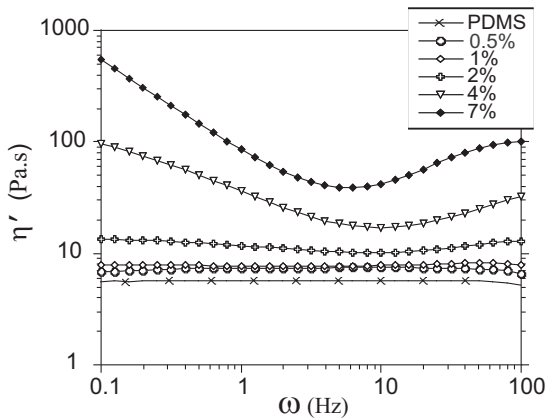}}
\caption{Summary for the dynamic viscosity $\eta'(\omega)$ against
frequency for well-dispersed samples of different concentrations.}
\label{wellmixA}
\end{figure}

\begin{figure} 
\centering
\resizebox{0.4\textwidth}{!}{\includegraphics{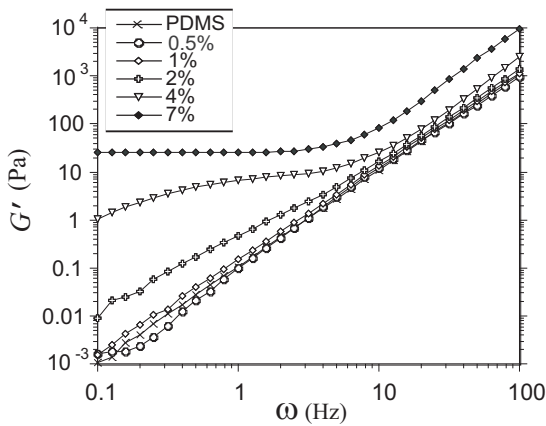}}
\caption{Summary for the storage modulus $G'(\omega)$ against
frequency for well-dispersed samples of different concentrations.
Note the emerging low-frequency rubber plateau in $G'$ at high
tube concentrations.} \label{wellmixB}
\end{figure}

The well dispersed states can be characterized by their
reproducible profile of the rheological linear response.
Figures~\ref{wellmixA} and \ref{wellmixB} give a summary of these
results, in both $\eta'$ and $G'$ representations. Increasing
nanotube concentration increases the values of $\eta'$, plot (a),
and also causes it to become more frequency dependent. The
0.5wt\%, 1wt\% and 2wt\% samples, just like the pristine PDMS,
exhibit a nearly frequency-independent Newtonian plateau in the
range of frequencies studied. These systems are dilute enough so
that the effect of hydrodynamic interaction between tubes is
negligible. There is a significant change in the viscosity
profiles between 2wt\% and 4wt\%, which suggests a major change in
nanocomposite structure. Note that these are the concentrations at
which we have verified the onset of nanotube interactions, cf.
Fig.~\ref{finalvisc}.

In order to identify the subsequent change in microstructure, we
plot the storage shear modulus $G'$ against frequency,
Fig.~\ref{wellmixB}. This is discussed in more detail in the next
section. Here we only wish to attract the attention to the
emerging rubber plateau, the static gel modulus $G'(\omega
\rightarrow 0)$ for highly interacting nanotube dispersions.

Both our $G'$ and $\eta'$ values (in the well-mixed state) are
$\sim$1 to 2 orders of magnitude lower for the same concentration
of nanotubes than the results in the
literature~\cite{Potschke2002,Song2005}. This is almost certainly
due to the fact that our PDMS matrix has lower initial viscosity
than the other systems investigated. However, in view of our
findings about the erratically high values of response moduli in
the state with insufficient tube dispersion (at $t<t^*$), one has
to be cautious about the details of preparation of polymer
nanocomposites: have the specific polymer / nanotube sample been
mixed for a sufficient time at a given shear stress of mixing?
Such a question is rarely addressed in the current literature,
making comparison difficult.

The change in rheological behavior as concentration of tubes
increases, similar to those presented in Figs.~\ref{wellmixA} and
\ref{wellmixB} have been reported for other nanotube (single- or
multi-walled) polymer composites and is often called the
`percolation threshold'~\cite{Potschke2002}. More precisely, one
might call the emergence of the static gel network the mechanical
percolation threshold, so as to differentiate it from the more
traditional electrical percolation~\cite{Du2004}. Again, there are
large discrepancies reported in the literature for such mechanical
percolation concentrations, even for the same system. A reason for
this might well arise because of two distinct possibilities.
Firstly, by forming a well dispersed and homogenous ($t_{\rm mix}
> t^*$) \textit{network} of nanotubes one may reach, and exceed,
the entanglement limit. In this case the rheological response
would become that of an elastic solid. Secondly, mechanical
percolation could take place when individual \emph{aggregates}, or
tube clusters (at $t_{\rm mix} < t^*$), come in contact and form
force chains. This second type of aggregate-mediated jamming may
well be responsible for much higher threshold concentrations and
moduli previously reported. Better dispersed samples of very long
nanotubes will naturally provide much lower percolation
thresholds.

\subsubsection{Mixing at high tube concentrations}\label{highwt}

As the weight fraction of nanotubes increases, the mixing
mechanism and the resulting dispersed structure also becomes more
complicated. For instance, for 7wt\% samples, empirical
observations suggest that the mixture was not continuous during
mixing: for the first few hours of mixing, clusters were sticking
to the walls of the container. Some clusters were then observed to
migrate to form larger structures (the whole process resembled the
mixing of a slurry). As the mixing time lengthened ($t_{\rm mix}>
10$ hours), large clusters `connected' back with the main mixture
and were eventually broken down.

The most significant change associated with the critical
concentration of nanotubes above which tube interactions are
relevant is the presence of an entangled elastic network structure
as the state of full dispersion is approached, $t \sim t^*$.
Evidence for this is shown by the rubber plateau in the storage
modulus at $\omega \rightarrow 0$ and also in the peaks in the
loss factor $\tan \delta$.   Because the relatively high stiffness
(large persistence length) of nanotubes, the amount of
entanglements that can form are a lot less compared to that of
common polymers (thus relatively low $G'$ of the network). On the
other hand, the high polarizability of the tubes also implies that
once these entanglements are formed, they are hard to disconnect.
All the tubes are connected by these physical crosslinks and form
a homogeneous network structure in the PDMS matrix. The rise in
the $\tan (\delta)$ peak with increasing mixing time is a useful
empirical method by which to confirm and quantify nanotube
dispersion.
\begin{figure}
\centering
\resizebox{0.4\textwidth}{!}{\includegraphics{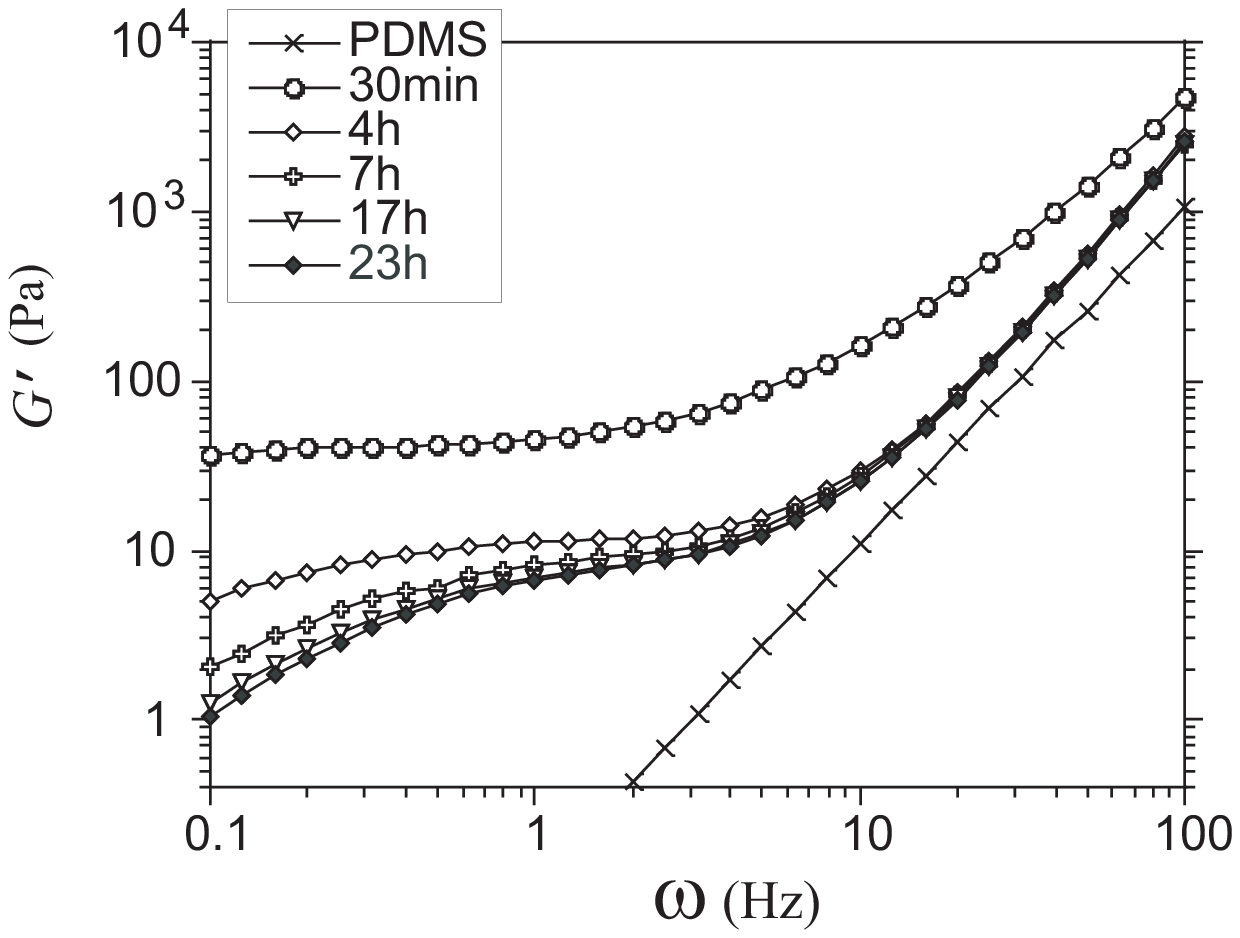}}
\caption{Data of $G'$ against frequency for 4wt\% samples mixed
for different times (tested immediately after mixing).
Non-representative profiles (similar to `cluster' curves in
Fig.~\ref{clustersA}) due to erratic behavior at $t<t^*=$16h have
been removed for clarity.} \label{4pcA}
\end{figure}

\begin{figure}
\centering
\resizebox{0.4\textwidth}{!}{\includegraphics{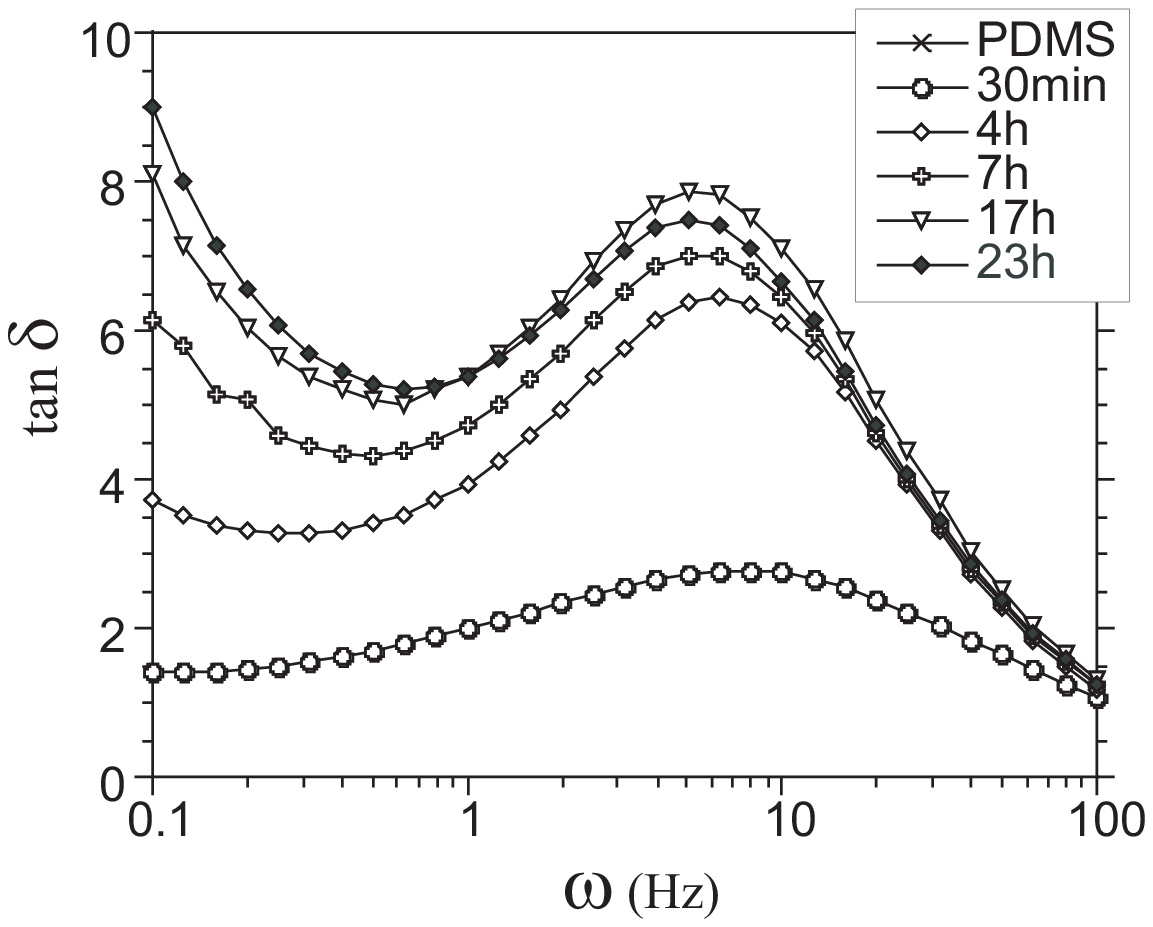}}
\caption{Data for $\tan \delta$ against frequency for 4wt\%
samples mixed for different times (tested immediately after
mixing). Non-representative profiles due to erratic behavior at
$t<t^*=$16h have been removed for clarity.} \label{4pcB}
\end{figure}

\begin{figure}
\centering
\resizebox{0.47\textwidth}{!}{\includegraphics{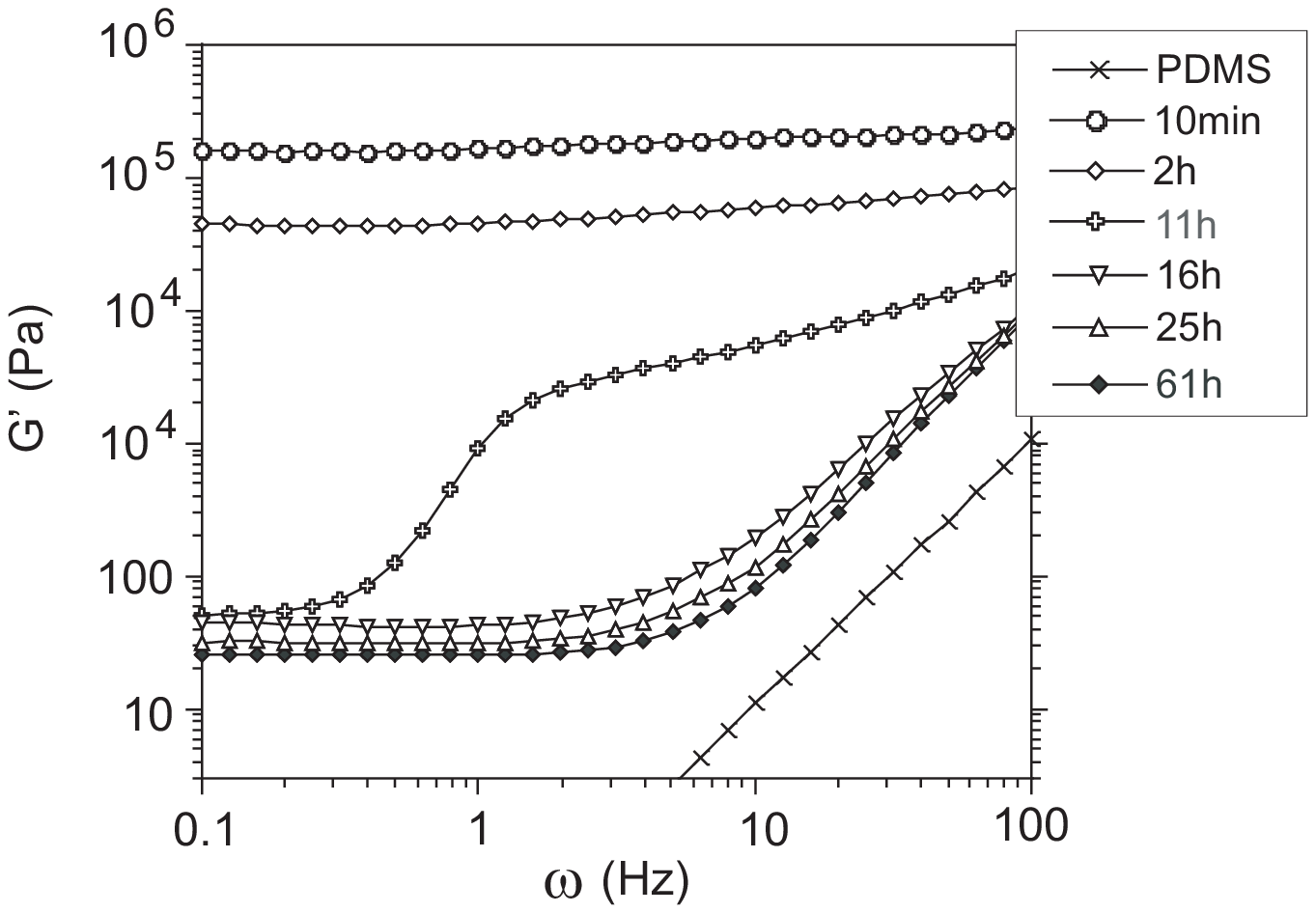}}
\caption{Data for $G'$ against frequency for 7wt\% samples mixed
for different times (tested immediately after mixing).
Non-representative profiles due to erratic behavior at $t<t^*=$23h
have been removed for clarity.} \label{7pcA}
\end{figure}

\begin{figure}
\centering
\resizebox{0.47\textwidth}{!}{\includegraphics{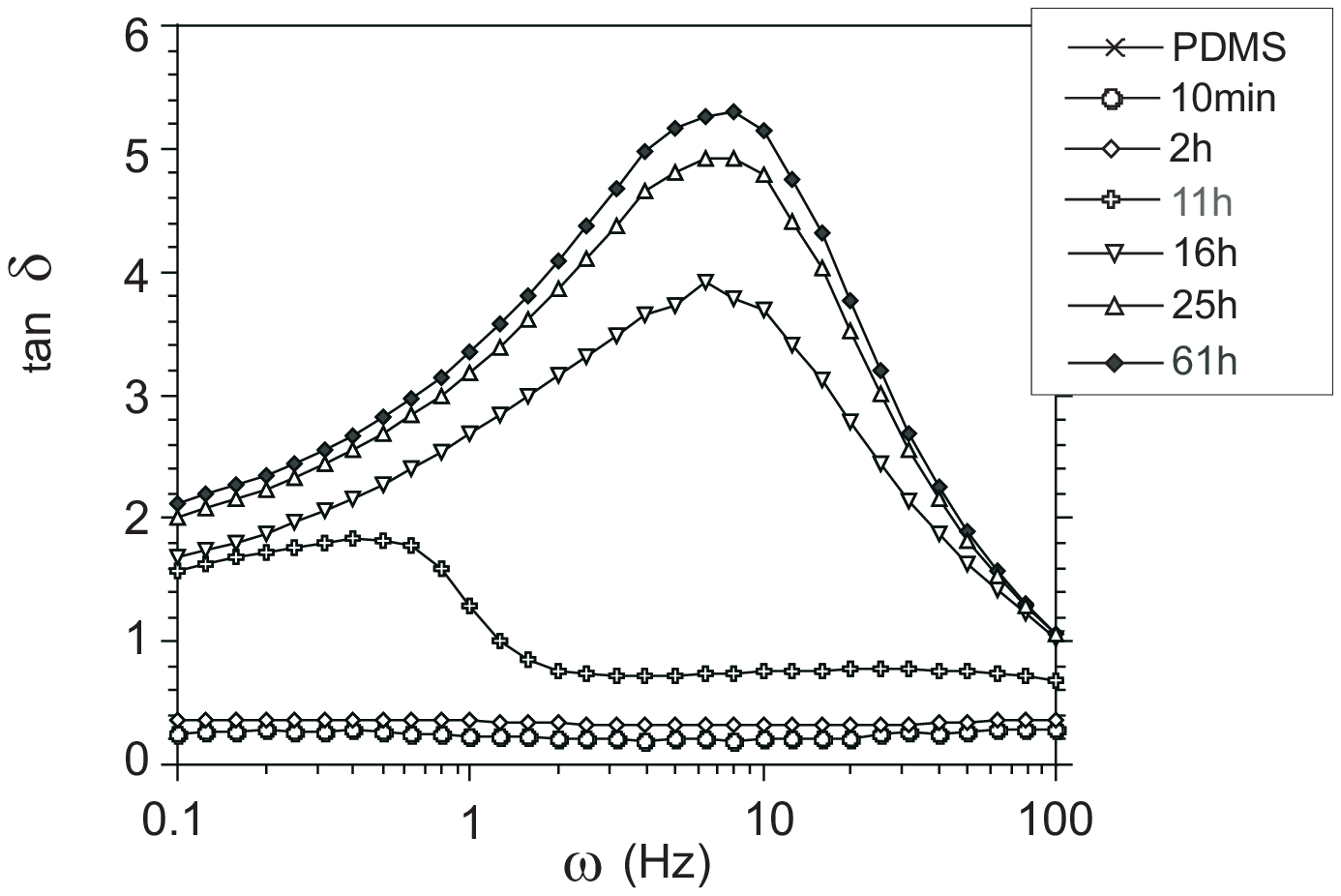}}
\caption{Data for $\tan \delta$  against frequency for 7wt\%
samples mixed for different times (tested immediately after
mixing). Non-representative profiles due to erratic behavior at
$t<t^*=$23h have been removed for clarity.} \label{7pcB}
\end{figure}

In general, the change of $G'(\omega)$ profile for the 4wt\%
composite, Figs.~\ref{4pcA} and \ref{4pcB}, is similar to that of
the 7wt\% sample. However, the latter has a more pronounced change
in its rheological response as the mixture homogenizes.
Figure~\ref{7pcA} shows a distinct difference in $G'(\omega)$
between the viscous PDMS liquid and the essentially solid nanotube
composite after just 10 minutes of mixing. We associate the high
plateau ($G'\sim 100$kPa) with a state of colloidal glass, or a
rheologically jammed state of tube clusters. As the time of mixing
increases, we observe the characteristic step of a glass
transition travel through the $G'(\omega)$ plots from the low to
the high frequencies. The end, in the well-mixed dispersion, is a
much lower rubber plateau ($G'\sim 20-40$Pa) of the homogeneous
entangled nanotube network. The corresponding movement of the
$\tan \delta$ peak in Fig.~\ref{7pcB} tells essentially the same
story.

\subsection{Re-aggregation and aging}\label{aging}

1wt\% and 7wt\% samples were selected to test the stability of
homogeneous dispersion, samples with nanotube concentration below
and above the critical entanglement value. The results for the
two-week aging of 1wt\% composite are in fact shown in
Fig.~\ref{threeruns}, back in Section 3.1. The sample mixed for
just 2min ($t_{\rm mix} \ll t^*$) we have seen in a large drop in
$\eta'$ upon standing for 2 weeks, essentially recovering a pure
PDMS value. In contrast, very little change in viscosity was
registered for the same aging of well-mixed samples, at $t_{\rm
mix}>t^*$. The apparent conclusion is that sparse non-interacting
nanotubes dispersed in a viscous matrix do not have enough
Brownian mobility to re-aggregate.

\begin{figure} 
\centering
\resizebox{0.49\textwidth}{!}{\includegraphics{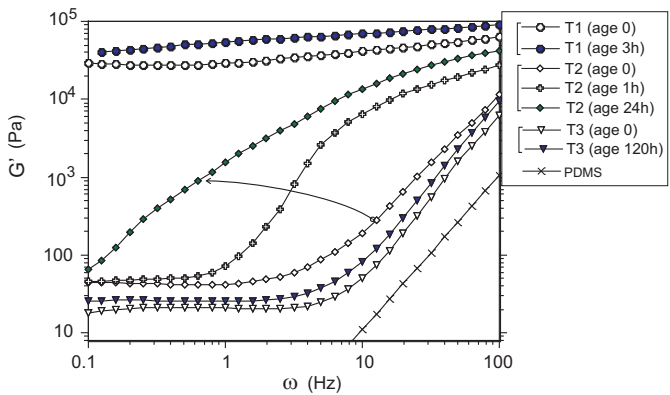}}
\caption{Plots of $G'$ against angular frequency for 7wt\%
samples. Three groups of data are labelled: {\sf T1}, for the
short mixing time $t\ll t^*$, {\sf T2}, for the intermediate state
of mixing ($t \leq t^*$), and {\sf T3}, for the well-dispersed
composite ($t \gg t^*$). The age of each sample is labelled in the
legend. } \label{7aging}
\end{figure}

Above the percolation concentration, the 7wt\% composite exhibits
a very different aging behavior. Figure~\ref{7aging} shows two
groups of measurements, for the samples mixed for a short time
($t_{\rm mix} =$1h, labelled T1 in the plot), for the intermediate
stage of mixing  ($t_{\rm mix}=$16h, labelled T2) and for the
well-dispersed composites ($t_{\rm mix}=$61h, labelled T3). The T1
response function $G'(\omega)$ shows only weak aging and remains
similar to the high value of colloidal glass modulus, cf. top
curves in Fig.~\ref{7pcA}. This observation is consistent with the
idea of a dynamical glassy state of jammed nanotube clusters.

The T2 response function $G'(\omega)$ evolves most significantly
with the aging time. This is highlighted by the arrow drawn on the
plot in Fig.~\ref{7aging}. The freshly made sample has the
response of an entangled elastic gel, with the plateau modulus at
$\omega \rightarrow 0$, as discussed above. Although this was not
yet a well-dispersed sample, the main features of tube
entanglement have already became apparent (cf. second from the
bottom $G'$ curve in Fig.~\ref{7pcA}). However, as this sample
ages, the rheological transition in this metastable gel moves to
the lower frequencies. In effect, we are observing the movement of
the dynamic glass transition across the plot in Fig.~\ref{7aging}.
It seems reasonable to assume that after a very long time the
modulus would revert to the colloid glass values characteristic of
the jammed cluster system T1.

Finally, the rheological signature of a well-dispersed sample T3
shows, once again, only a very small aging in spite of a very long
time allowed for this sample to stand (6 days). The rubber plateau
modulus $G' (\omega \rightarrow 0) \sim $20~Pa remains almost
unchanged, as does the frequency of the transition. We may
conclude that the entangled network of nanotubes, with no clusters
left to act as seeds for re-aggregation, represents a very deep
energy well of a dispersion metastable state.

In order to explain the phenomena observed in the standing samples
with low concentrations, both sedimentation and re-aggregation
processes should be considered. For spherical particles, the ratio
of gravitational to Brownian forces $a^{4}\Delta \rho g/(k_{B}T)$
should be less than unity so as to avoid sedimentation. Applying
this idea to our system, with the density difference between
compacted nanotubes and the PDMS of order $\Delta \rho \sim 0.6
\hbox{g/cm}^3$, it shows that any clusters of sizes greater than
$\sim 1.5\mu$m would tend to settle at room temperatures. The high
viscosity of the medium may require a long time for the effect to
become noticeable, especially for smaller sized clusters.

By comparing Figs.~\ref{7aging} and \ref{7pcA}, it is clear that
the relaxation, or aging, is a reverse process of dispersion, the
sample T2 clearly being dispersed into a metastable state prone to
relaxation. Further development of an entangled elastic network
would serve to stabilize the homogeneous dispersed state, so that
no substantial relaxation is observed within the duration of our
experiment -- but it is nevertheless going to happen in the same
way.

The implication of the dispersion aging is important to note.
Firstly, even an ideal homogeneous dispersion achieved when
$t_{\rm mix} \geq t^*$ may well disappear after a time, unless
further processing steps are taken to `freeze-in' the dispersed
structure, e.g. by chemical cross-linking of the polymer matrix.
Storage conditions should be changed so as to minimize the
alteration, e.g. by cooling the matrix below the polymer glass
transition. Secondly, in order to use rheological data to study
the effect of mixing time on dispersion, special caution should be
taken for the higher loaded samples ($> 2$\,wt\% in our case) due
to the relaxation processes that are sensitive to standing times.

\section{Conclusion}
Experiments have shown that a critical time $t^*$ is needed to
disperse carbon nanotubes in a polymer melt, reaching a consistent
and reproducible state of such a dispersion. Below this
characteristic time, the composite system is full of dense tube
clusters (often smaller than an optical microscope resolution).
This manifests itself in erratic rheological properties, depending
on accidental jamming of the resulting ``colloidal glass'' system.
Dispersions mixed for a time longer than $t^*$ appear
homogeneously mixed, with their rheological behavior shows no
evidence of jamming. We may only hope a complete dispersion has
been achieved at mixing times above $t^*$. One cannot exclude a
presence of consistently small tube clusters or bundles, and there
is no unambiguous technique to confirm or disprove this. However,
a homogeneous dispersion is suggested by images of
freeze-fractured surfaces, Fig.~\ref{mixer}(b), and by comparing
the estimates of semiflexible overlap and entanglement
concentrations with our rheological measurements. For all
practical purposes we regard our composite at $t>t^*$ as
completely dispersed, which is hardly the case in the present
literature.

The critical time of mixing, $t^*$, is a function of nanotube
concentration and the shear stress in the mixing device (itself a
function of vessel geometry and the viscosity of the polymer
matrix). Although we have not done such a study, it is quite
obvious that the shear stress energy delivered to the particles
during mixing ($\sim 10^6\hbox{J/m}^3$ in our case) has to exceed
the van der Waals force of their attraction in the contact region
(crudely estimated as, $\sim 10^4\hbox{J/m}^3$, from
\cite{Broni2001}). Below this shear energy density one cannot hope
to achieve dispersion no matter how long is the mixing.

The second important parameter in nanotube dispersion is their
concentration. Well-dispersed systems possess very different
rheological properties below and above the concentration of
``mechanical percolation'' (we use this term reluctantly, only
because it seems to be in heavy use in the literature: the true
percolation is a somewhat different physical process). At low
concentrations, non-interacting nanotubes homogeneously dispersed
in the polymer matrix appear quite stable against re-aggregation
(provided the matrix viscosity is high enough to suppress fast
Brownian motion). The rheology of such dispersions remains that of
a viscous liquid, with the observed steady-state viscosity a
linear function of tube concentration (the slope of this
dependence, much higher than the classical Einstein's 2.5, is
certainly due to the profound shape anisotropy of nanotubes).

At concentrations above the threshold of order 2-3wt\% in our case
we see a clear emergence of an elastic gel of entangled nanotubes
in their homogeneously dispersed state. This agrees favorably with
an estimate of overlap concentration $n_c \sim 1.5$\,wt\% made in
Section~II (considering the inevitably crude nature of  $n_c$
estimate). The rheological characteristics of these composites
have a distinct rubber modulus $G'$ in the limit of zero
frequency. We also note a characteristic superposition between the
mixing time and the frequency of rheological testing, similar to
the time/temperature superposition in classical glass-forming
polymers. Here we find the transition between the colloidal glass
state of jammed tube clusters at short mixing times, and the
weakly elastic state of an entangled nanotube gel.

We hope these results, as well as the brief study of
re-aggregation and sedimentation of dispersed polymer
nanocomposites, would contribute to a more rigorous and
quantitative approach of preparation and analysis of carbon
nanotube dispersions.

\subsection*{Acknowledgements} We thank S.F. Edwards, A. Craig and
A.R. Tajbakhsh for insightful discussions. This work was carried
out with the support of the EPSRC, the ESA-ESTEC (18351/04) and a
CASE award from Makevale Ltd.

\end{document}